\documentclass[twocolumn,showpacs,amsmath,amssymb]{revtex4}
\usepackage{graphicx} 
\usepackage{mathrsfs}

\newcommand{\be}{\begin{eqnarray}}
\newcommand{\ee}{\end{eqnarray}}
\newcommand{\la}{\langle}
\newcommand{\ra}{\rangle}

\newcommand{\gp}{g^\prime}
\newcommand{\gt}{\sqrt{g^{\prime2}-g^2}}

\newcommand{\Tr}{{\rm Tr}}
\begin{document}
\title{Black Holes Are Almost Optimal Quantum Cloners}

\author{Christoph Adami$^{1}$ and Greg  ver\! Steeg$^2$\\}
\mbox{}\vskip 0.5cm \affiliation{
$^1$Department of Physics and Astronomy, Michigan State University, East Lansing, MI 48824\\
$^2$Information Science Institute, Viterbi School of Engineering, University of Southern California, Los Angeles, CA 90089\\
}
\date{\today}
\begin{abstract}If black holes were able to clone quantum states, a number of paradoxes in black hole physics would disappear. However, the linearity of quantum mechanics forbids exact cloning of quantum states. Here we show that black holes indeed clone incoming quantum states with a fidelity that depends on 
the black hole's absorption coefficient, without violating the no-cloning theorem because the clones are only approximate. Perfectly reflecting black holes are optimal universal ``quantum cloning machines"
and operate on the principle of stimulated emission, exactly as their quantum optical counterparts. In the limit of perfect absorption, the fidelity of clones is only equal to what can be obtained via quantum state estimation methods. But for any absorption probability less than one, the cloning fidelity is nearly optimal as long as $\omega/T\geq10$, a common parameter for modest-sized black holes. 
\end{abstract}
\maketitle

Black holes are quantum objects with intriguing characteristics. They are formed in the stellar collapse of stars with sufficient mass because such objects become relativistically unstable. While classically no signal can emerge from within the event horizon, Hawking showed that in curved-space quantum field theory, black holes must emit thermal radiation with a temperature $T$ that is inversely proportional to the black hole's mass $M$: $T=1/(8\pi M)$, in convenient units~\cite{Hawking1975}. 

But Hawking's calculation also created an apparent paradox. Because the eponymous radiation takes its energy from the mass of the black hole, the latter might ultimately evaporate. And if the emitted radiation is strictly thermal as the calculation suggests, the initial data about the formation of the black hole would have to be erased once the black hole disappears, something the laws of physics simply should not permit~\cite{Witten2012}. The same appears to hold true for information directed at the event horizon after the formation of the black hole (this means ``at late times" in the parlance of gravity, because according to stationary observers the black hole only forms for $t\to\infty$). If the black hole final state does not depend on whether Shakespeare's or Goethe's works are absorbed by the black hole--assuming that both {\em \oe uvres} have the same mass--then  space time dynamics would appear to be irreversible and unpredictable even if black holes never evaporate at all. 

Recently, we found that from the point of view of quantum communication theory, black holes are not so special after all~\cite{AdamiVersteeg2014}. Rather, they appear to be relatively ordinary noisy quantum communication channels, with a positive capacity to transmit classical information given by the 
expression proposed by Holevo~\cite{Holevo1998}, because any matter or radiation absorbed by the black hole must stimulate the emission of {\em exact copies} outside of the event horizon~\cite{BekensteinMeisels1977}. Because the absorbed particles are quantum states, we may wonder whether stimulated emission at the black hole horizon violates the quantum no-cloning theorem~\cite{Dieks1982,WoottersZurek1982}. In the following we show that it does not, because the spontaneously emitted Hawking radiation prevents perfect cloning. We will ask how well quantum states are copied by black holes, because this will shed light on how black holes treat {\em quantum}--rather than classical--signal states. We find that a black hole's cloning fidelity depends on the probability with which quantum states are absorbed by the black hole, $\Gamma_0$, with a cloning fidelity ranging from the classical measurement limit to the universal optimal cloner~\cite{BuzekHillery1996,GisinMassar1997}.

\begin{figure}[htbp]
\begin{center}
\includegraphics[width= 5.25cm,angle=0]{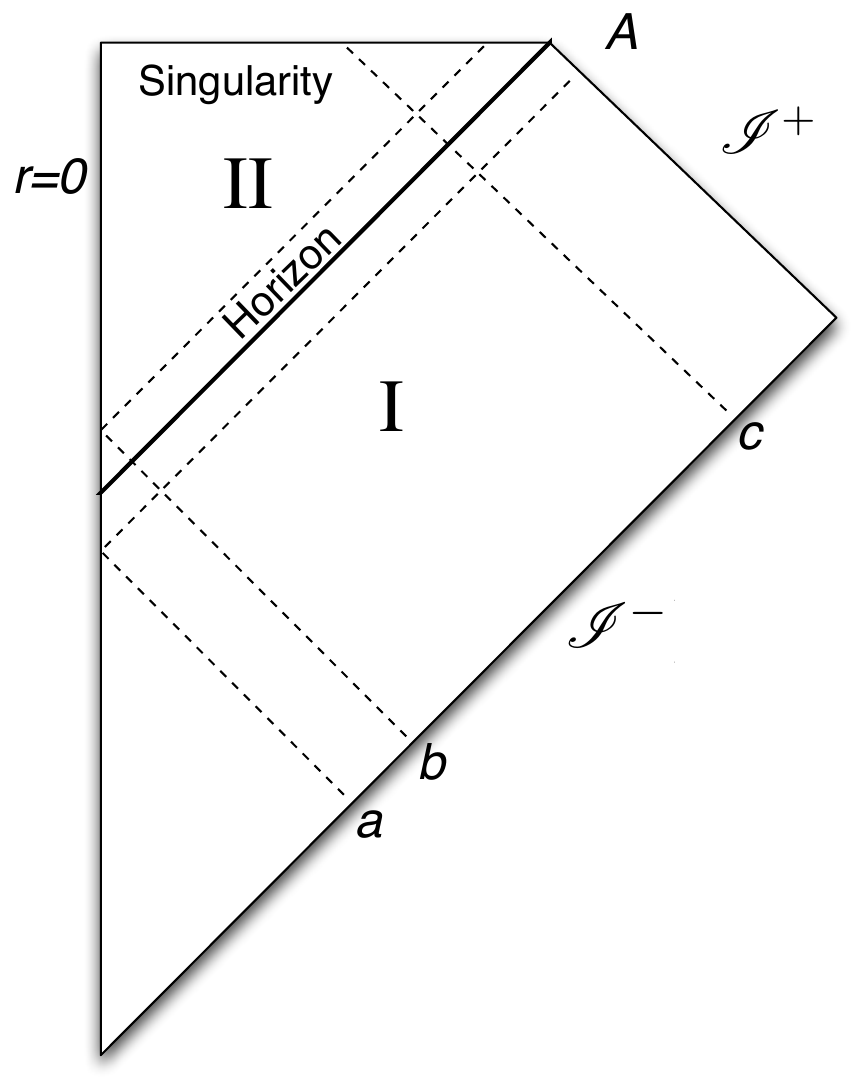}
\caption{Penrose diagram of the spacetime of a black hole. Modes $a$, $b$,
$c$ and $A$ are concentrated in a region of null infinity
indicated by the letter (note that $a$ and $b$ actually overlap on
${\mathscr I^-}$, and that our nomenclature of modes differs from that of Sorkin~\cite{Sorkin1987})  } \label{default}
\end{center}
\end{figure}

Quantum black holes can be described succinctly by a hermitian operator $H$ that transforms 
quantum states defined on a Fock space of particles {\em before} the formation of the black hole and far away from the star that will form it (a slice of spacetime called ``past null infinity" and denoted by ${\mathscr I^-}$  in Fig. 1), to a Fock space long after formation of the black hole, also far away from it (``future null infinity" ${\mathscr I^+}$ in Fig.~1). As opposed to flat-space quantum field theory where Poincar\'e invariance ensures that those two vacua are equivalent, curved-space quantum field theory can have inequivalent vacua defined only by specifying a time-like coordinate with respect to which field modes carry positive frequency (see, e.g.,~\cite{BirrellDavies1982}). 

We will define our ``in-vacuum" by separating space-time into two regions divided by an event horizon: the ``inside" region (denoted by ``II'' in Fig.~1) and an ``outside" (region I). For each region, we can define a complete set of modes that in the future travel just outside and just inside the event horizon, and creation and annihilation operators ($a_k$ annihilates modes in the outside region whereas $b_k$ annihilates modes on II). Because the in-modes so defined diverge on the horizon, a different set of coordinates is needed to describe the out-states. These so-called ``Kruskal" modes are regular on the horizon, and related to the in-states via
\be
|\psi_{\rm out}\ra= e^{-iH}|\psi_{\rm in}\ra\;. \label{trans}
\ee
where $H$ is a Hermitian operator defined below.

Associated with these out-states are creation and annihilation operators $A_k$ (annihilating outside states defined on ${\mathscr I^+}$) and $B_k$ (annihilating ``inside" states defined on the horizon). The change of coordinate systems implied by Eq.~(\ref{trans})
gives rise to a  Bogoliubov transformation relating $A_k$ to $a_k$ and $b_k$:
 \be
A_k &=& e^{-iH}a_k e^{iH} = \alpha_k a_k - \beta_k b^\dagger_{-k}\label{bog1}\;,
\ee
which can be implemented by the Hermitian operator
\be
H = ig_k(a_k^\dagger b_{-k}^\dagger - a_k b_{-k} + k\rightarrow
-k)  \label{ham}\ee
describing the entanglement of particles
and antiparticles outside and inside the horizon (for mode $k$). For a theory of complex scalar fields in curved space-time, we find 
 \be
\alpha_k^2 &=& \cosh^2 g_k = \frac1{1-e^{-\omega/T}}\;\\
\beta_k^2 & = & \sinh^2 g_k = \frac1{e^{\omega/T} -1}\;, 
\ee 
where
$\omega=|k|$ is the frequency associated with mode $k$ and $T$ is
the Hawking temperature. The relationship between the coefficients $\alpha_k$, $\beta_k$, the frequency of the modes and the Hawking temperature $T$ is enforced by  requiring that the solution to the free-field equations in a black hole background is analytic across the horizon, as usual~\cite{Hawking1975}. Note that while the full Hamiltonian is a sum over modes with all $k$, we focus here on a single mode because they do not mix in this 1+1 dimensional theory. Note further that while we could have restricted ourselves to real-valued scalar fields, the ability to distinguish particles from antiparticles will be useful later when we encode quantum states.

Quantum cloning refers to the duplication or copying of quantum information, a process that is known to be impossible to achieve perfectly~\cite{Dieks1982,WoottersZurek1982}
but that can be performed approximately. A quantum cloning machine attempts to maximize the probability that a quantum state $|\psi\rangle$ (or by extension $N$ identically prepared states $|\psi\ra^{\otimes N}$) is cloned into a state $|\psi_{\rm out}\ra=U|\psi_{\rm in}\rangle|0\ra |X\ra$ with a unitary transformation $U$ acting on the input state $|\psi_{\rm in}\ra$, a blank state $|0\ra$ that is to hold multiples of the cloned state, and an ancillar state $|X\ra$. The formalism of quantum cloning machines was introduced by Buzek and Hillery~\cite{BuzekHillery1996}, and has received considerable attention (see, e.g., the reviews~\cite{CerfFiurasek2005,Scaranietal2005}). The fidelity of one of the $M$ cloned states $\rho_{\rm out}^j$ is given by
\be
F_j=\mbox{}_{\rm in}\la\psi|\rho^j_{\rm out}|\psi\ra_{\rm in}\;,
\ee
and the optimal $N\to M$ cloning fidelity of universal quantum cloning machines (devices that copy any input state $|\psi\ra$ with the same fidelity) is~\cite{GisinMassar1997,Brussetal1998}
\be
F_{\rm opt}=\frac{M(N+1)+N}{M(N+2)}\;. \label{optfidel}
\ee
Quantum cloning machines that reach the optimal cloning limit have been constructed using simple quantum optical elements (parametric amplifiers and beam splitters) for discrete states such as polarization~\cite{Simon2000} as well as continuous variables~\cite{Fiurasek2001,Braunsteinetal2001}.

In order to discuss quantum cloning by black holes, we study the effect of the unitary transformation $U=e^{-iH}$ applied to an arbitrary incoming quantum state $|\psi_{\rm in}\ra$ (including the blank and ancillar states), where $|\psi\ra$ encodes information in the ``particle-antiparticle" basis, i.e., $|\psi_{\rm in}\ra=\sigma|1,0\ra_g+\tau|0,1\ra_g$ and where $|n,m\ra_g$ represents a state with $n$ particles and $m$ antiparticles at past null infinity (the time slice ${\mathscr I^-}$ in Fig.~1).

Let us first discuss cloning with the operator Eq.~(\ref{ham}), because it allows us to establish important limits. The modes we use to encode our logical qubits
are number states of particles or antiparticles, timed in such a manner that they travel just outside (modes $a_k$) or just inside (modes $b_k$) the event horizon. Naturally, this is not a practical means of cloning quantum states because particles cannot be timed this accurately if the formation of a black hole is not certain. Furthermore, particles traveling close to the horizon for long times undergo a tremendous redshift (which we neglect here for simplicity). Consequently, we will discuss cloning of {\em late-time} particles that are sent in long after the formation of the black hole (and that do not undergo a redshift) following this discussion.

For the unitary mapping $U=e^{-iH}$  implemented by (\ref{ham}),                    
we find that the cloning transformation  is rotationally invariant just as in quantum optics~\cite{Simon2000}, implying that the cloner is {\em universal} (that is, that the cloning operation is independent of the state to which it is applied). Thus, instead of considering the effect of the cloner on the general state $|\psi\ra=\sigma|1\ra_L+\tau|0\ra_L$, we can choose $\sigma=1$ and $\tau=0$  without loss of generality. 

For $|\psi_{\rm in}\ra = |N,0\ra_a|0,0\ra_b=|1\ra_L^{\otimes N}$ we obtain using (\ref{ham})
\be
|\psi_{\rm out}\ra=Z\sum_{jj^\prime}^{\infty}e^{-(j+j^\prime)\frac{\omega}{2T}}\sqrt{\binom{j+N}{N}}|j+N,j^\prime\ra_a|j^\prime,j\ra_b\;, \nonumber \\
\ee
where $Z=(1/\alpha_k)^{2+n}$. Fixing the number of output copies to $M$ (by restricting the number of particles and antiparticles in region I to $M$) gives
\be
|\psi\ra_M\sim \sum_{j=0}^{M-N}(-1)^j \sqrt{\binom{M-j}{N}}|M-j,j\ra_a|j,M-N-j\ra_b\;, \nonumber\\ \!\!\! \label{post}
\ee 
which (up to normalization) is identical to the wavefunction emanating from the optical quantum cloner of Simon et al.~\cite{Simon2000} that achieves the optimal fidelity~(\ref{optfidel}). Thus, for quantum states sent into a black hole at early times so that they remain just outside the horizon, the black hole acts like a universal, optimal, quantum cloning machine~\footnote{In quantum optics, post-selection to a fixed number of clones can be achieved by entangling the quantum state with a trigger signal and then measuring the trigger, which reduces the wavefunction to Eq.~(\ref{post}) if $M$ particles were detected in the trigger. In the case of black holes, we can perform the same operation, but must send the trigger towards future infinity (but away from the black hole horizon). Even though the quantum state reconstruction is conditioned on the trigger, it only uses the quantum states coming from the black hole.}.

Let us study the quantum states {\em behind} the horizon (modes $b$). Just as in the optical realization of the optimal universal cloner, the quantum state behind the horizon contains $M-N$ {\it anticlones} of the initial state $|N,0\ra_a$, that is, inverted states 
obtained from the initial states by the application of the optimal universal NOT gate~\cite{Buzeketal1999,GisinPopescu1999}. The fidelity of these anticlones is
\be
F_{\rm anti}=\frac{N+1}{N+2}
\ee
for each anticlone, which is the fidelity of the best possible state preparation via {\em state estimation} from a finite quantum ensemble~\cite{MassarPopescu1995}. Note that the optimal cloning fidelity (\ref{optfidel}) tends to the state estimation fidelity in the limit $M\rightarrow\infty$, which we recognize as the classical measurement limit of quantum cloning.

Consider the fidelity of cloning when $N$ antiparticles in mode $b$ are sent in, traveling just inside the horizon (region II): $|\psi\ra_{\rm in}=|0,0\ra_a|0,N\ra_b$. We are now interested in the fidelity of the anticlones generated on the other side (region I). The wavefunction is
\be
|\psi\ra_M\sim \sum_{j=0}^{M}\sqrt{\binom{j+N}{N}}|j,M-j\ra_a|M-j,j+N\ra_b
\ee
giving the probability to observe $j$ particles and $M-j$ antiparticles outside the horizon
\be
p(j,M-j|N)=\frac{\binom{j+N}{N}}{\sum_{j=0}^{M}\binom{j+N}{N}}\;.
\ee
The fidelity of these anticlones in region I is also $N+1/N+2$, equal to the fidelity of the anticlones in II when sending in $|N,0\ra_a$, but the anticlones of the antiparticle states are of course clones of the particle states. In summary, sending $N$ particles in mode $a$ creates optimal particle clones in region I, while sending in $N$ antiparticles in mode $b$ gives rise to classical particle clones in region I instead.

We now discuss the more realistic cloning scenario where we send quantum states into the already-formed black hole at late times, by introducing an additional mode $c$ (see Fig.~1) that is strongly blue-shifted with respect to the early-time modes $a$ and $b$ (which have suffered a redshift from traveling through a dynamic collapsing space time), and therefore commutes with them. The $c$-modes are turned into $a$ and $b$ modes via absorption, reflection, and emission at the horizon, and were used earlier as signal states to study the fate of information sent into black holes~\cite{AdamiVersteeg2014}.
The Bogoliubov transformation that connects these operators (for each mode $k$) is~\cite{Sorkin1987} (we omit the index $k$ from the Bogoliubov coefficients from now on)
\be
A_k=e^{-iH}a_ke^{iH}=\alpha a_k-\beta b^\dagger_{-k}+\gamma c_k\;, \label{bog}
\ee
which can be implemented with the Hamiltonian
\be
H = ig(a_k^\dagger b_{-k}^\dagger - a_k b_{-k} )+ig^\prime (a_k^\dagger c_k- a_k c_k^\dagger) + k\rightarrow -k\;.\ \ \label{ham2}
 \ee
This Hamiltonian has an interesting quantum optical interpretation as the sum of an active optical element--a parametric amplifier described by a squeezing Hamiltonian with gain $g$--and a passive one: a beam-splitter with a phase $g^\prime$ (see, e.g.,~\cite{LeonhardtNeumaier2004}). It describes a black body at temperature $T$ surrounded by a semi-transparent mirror, which models the potential barrier surrounding the black hole~\cite{Buzek1998}.
 
Obtaining the coefficients $\alpha,\beta$, and $\gamma$ from (\ref{ham2}) is straightforward, and yields~\cite{AdamiVersteeg2014}  
 \be
\alpha^2 &=& \cos^2(\gt)=\Gamma_0=\frac{\Gamma}{1-e^{-\omega/T}}\; \label{alpha}\\
\beta^2&=&(\frac{g}{\gp})^2\frac{\sin^2(\gt)}{w^2}=\frac{\Gamma}{e^{\omega/T}-1}\; \label{beta}\\
\gamma^2&=&\sin^2({\gt})/w^2=1-\Gamma\;. \label{gamma}\ee where
$w^2=1-(g/\gp)^2$. Note that unitarity implies $\alpha^2-\beta^2+\gamma^2=1$. The coefficients $\alpha$ and $\beta$ together again set the black hole temperature via $\beta^2/\alpha^2=e^{-\omega/T}$. The quantum absorption probability of the black hole, $\Gamma_0$, is set by $\alpha^2$ alone, while the classical absorption probability $\Gamma$ is set by $\alpha^2-\beta^2$. Thus, $\Gamma$ is always smaller than $\Gamma_0$ by $\beta^2$, which is the strength of the quantum effects. In the above formulas, $g^\prime\geq g$ so that $w$ is real and $0\leq\Gamma_0\leq1$~\cite{BekensteinMeisels1977,Schiffer1993} is a probability~\cite{AdamiVersteeg2014}.

Consider first $1\rightarrow M$ cloning. Because the extended Hamiltonian (\ref{ham2}) is still rotationally invariant, we can again restrict ourselves to study the effect of the black hole on one particular state. To clone the state $|1\ra_L=|0,0\ra_a|0,0\ra_b|1,0\ra_c$, for example, we obtain: 
\be
\rho_a=\Tr_{bc}\left(U|1\ra_L\la1|U^\dagger\right) = \rho_{k|1}\otimes\rho_{-k|0}\;,
\ee
where (with  $\xi=\frac{\gamma^2}{\alpha^2\beta^2}$)
\be
\rho_{k|1}&=&\sum_{m}p(m|1)|m\ra\la m| \nonumber\\
&=&\frac{\alpha^2}{(1+\beta^2)^2}\sum_{m=0}^\infty
\left(\frac{\beta^2}{1+\beta^2}\right)^{m}\!\!\!\!(1+m\xi) |m\ra\la m|\;, \ \ \ \ \label{rho1}\\
\rho_{k|0}&=&\frac1{1+\beta^2}\sum_{m=0}^\infty
\left(\frac{\beta^2}{1+\beta^2}\right)^{m} |m\ra\la m|\;.
\label{rho0}
\ee

\begin{figure}[thbp]
\begin{center}
\includegraphics[width= 8cm,angle=0]{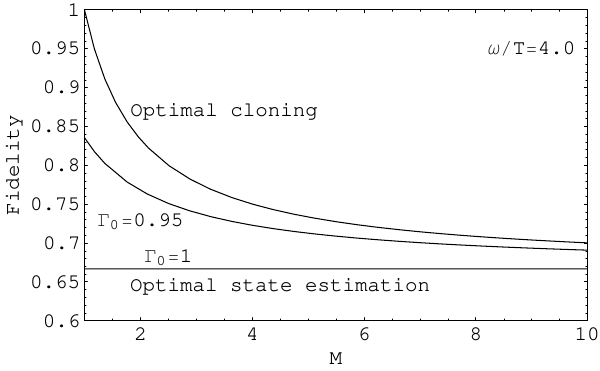}
\caption{Cloning fidelity $F_{1\to M}$ of the quantum black hole as a function of the number of copies $M$, for different values of the quantum absorption probability $\Gamma_0$  and a fixed $\omega/T=4$.   
\label{fig-fid2}}
\end{center}
\end{figure}

From (\ref{rho1},\ref{rho0}) the $1\rightarrow M$  cloning fidelity can be calculated as before 
\be
F_{1\rightarrow M}=\frac{\sum_{j=0}^{M}\frac{M-j}M p(M-j|1)p(j|0)}
{\sum_{j=0}^Mp(M-j|1)p(j|0)}=\frac{3+\xi+2\xi M}{3(2+\xi M)}\;.\ \ \label{bhfidel}
\ee
Let us investigate this result in a number of physical limits. As the black hole becomes more and more reflective, $\Gamma_0=\alpha^2\rightarrow0$, which implies $\xi\rightarrow\infty$.
In this case, the fidelity (\ref{bhfidel}) approaches the optimal value
\be
\lim_{\Gamma_0\rightarrow0} F_{1\rightarrow M}=\frac23+\frac1{3M}\;,\label{optfidel1}
\ee
as is seen in Fig.~\ref{fig-fid2}. For arbitrary $N$, the limit is exactly equal to the Gisin-Massar optimal fidelity (\ref{optfidel}) of an $N\to M$ cloning machine. We can recognize this result as the special case we treated earlier: If the black hole perfectly reflects incoming states, the black hole behaves just as if early-time modes ($a$-modes) were traveling just outside the horizon (except for the redshift).

Another limit of note is that of full absorption: $\Gamma_0\rightarrow1$. In that case $\xi\to1$ and $F_{1\to M}\to 2/3$, the fidelity of a classical cloning machine. It can be shown in general that for full absorption, the $N\rightarrow M$ cloning fidelity is equal to $N+1/N+2$ independently of $\omega/T$, which is the result we obtained earlier when sending $N$ antiparticles in mode $b$ directly behind the horizon. This is again not surprising, as the absorption of $c$-modes stimulates the emission of $b$ anti-modes behind the horizon, who in turn give rise to anticlones of the antiparticles, that is, clones. 

While in the limit $\Gamma_0\rightarrow1$  the best we can do to reconstruct the quantum state is to make classical measurements that allow us to optimally estimate the quantum state, note that in the limit $N\to\infty$ the probability to do this correctly tends to one, implying that the quantum state information can be reconstructed with arbitrary accuracy. 
\begin{figure}[tbp]
\begin{center}
\includegraphics[width= 8cm,angle=0]{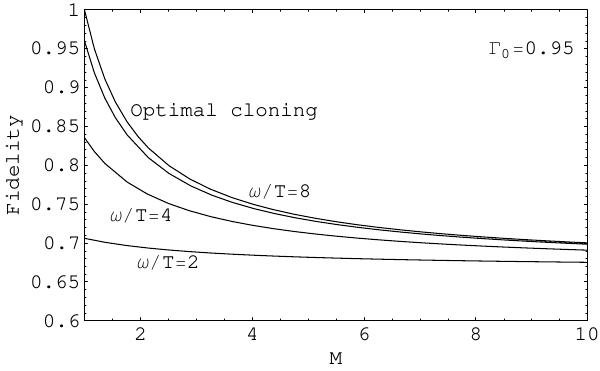}
\caption{Cloning fidelity $F_{1\to M}$ as a function of the number of copies $M$, for different values of $\omega/T$ and a fixed quantum absorption probability $\Gamma_0=0.95$.
\label{fig-fid3}}
\end{center}
\end{figure}
Another interesting limit is that of large black holes, where the Hawking temperature approaches zero. In that case, as  $\omega/T\to\infty$ implies $\xi\to\infty$ (as long as $\Gamma_0<1$),  we again recover the optimal universal quantum cloning fidelity (\ref{optfidel1}). This behavior can be seen in Fig.~\ref{fig-fid3}.
Note that as even modest-sized black holes have $\omega/T\geq10$, most black holes are nearly-optimal universal quantum cloners unless the absorption probability is exactly equal to 1. These results extend to $N\to M$ cloning machines. Just as for the $N=1$ case, the black hole cloner approaches the optimal cloner in the limit $T\to0$ or $\Gamma_0\to0$, and turns into a classical cloning machine in the limit $\Gamma_0\to1$ and for $M\to\infty$.

Our results have important implications for quantum communication theory, as the quantum channel defined by the mapping (\ref{trans}) turns out to be an example of so-called ``cloning channels"~\cite{Bradler2011}, whose classical capacity can be calculated exactly.  More precisely, the black hole acts a weighted ensemble of cloning machines. The quantum capacity of this channel has been calculated exactly in the limiting cases $\Gamma_0=0$ and $\Gamma_0=1$~\cite{BradlerAdami2014}. 

Our findings should have significant consequences for contemporary controversies surrounding black hole dynamics, such as black hole complementarity~\cite{tHooft1985,Susskindetal1993,Stephensetal1994} and firewalls~\cite{Almheirietal2012,Braunsteinetal2013}. Firewalls, that is, a ``curtain" of high-energy quanta surrounding a black hole, were proposed as a solution to an apparent inconsistency in back hole complementarity, which attempts to resolve the information paradox via quantum cloning (so that quantum states are both absorbed and reflected at the horizon). We have shown here that quantum states are indeed both reflected and absorbed, but that the dynamics occur in a unitary manner without violating the no-cloning theorem. Because the quantum state across the horizon is fully entangled in our description, an explicit calculation of the energy-momentum tensor must be finite across the horizon (while the energy-momentum tensor diverges if a product state across the horizon is used, as it assumed in the AMPS construction). An explicit calculation of the renormalized energy-momentum tensor~\cite{Davies1978}, in the presence of stimulated emission accompanying Alice's descent into the black hole, should resolve the paradox once and for all. 



%
\vskip 0.25cm \noindent{\bf Acknowledgements}
We are grateful to K. Br\'adler, Nicolas Cerf, and Paul Davies for discussions. This work was 
supported in part by the Army Research Office's grant \# DAAD19-03-1-0207.

\begin{thebibliography}{10}

\bibitem{Hawking1975}
S.~W. Hawking, ``Particle creation by black holes,'' {\em Commun. Math. Phys.},
  vol.~43, pp.~199--220, 1975.

\bibitem{Witten2012}
E.~Witten, ``Quantum mechanics of black holes,'' {\em Science}, vol.~337,
  pp.~538--40, 2012.

\bibitem{AdamiVersteeg2014}
C.~Adami and G.~{Ver Steeg}, ``Classical information transmission capacity of
  quantum black holes,'' {\em Class. Quantum Grav.}, vol.~31, p.~075015, 2014.

\bibitem{Holevo1998}
A.~Holevo, ``The capacity of quantum channel with general signal states,'' {\em
  IEEE Trans. Info. Theor.}, vol.~44, pp.~269--273, 1998.

\bibitem{BekensteinMeisels1977}
J.~D. Bekenstein and A.~Meisels, ``Einstein {A and B} coefficients for a black
  hole,'' {\em Phys. Rev. D}, vol.~15, p.~2775, 1977.

\bibitem{Dieks1982}
D.~Dieks, ``Communication by epr devices,'' {\em Phys. Lett.A}, vol.~92,
  p.~271, 1982.

\bibitem{WoottersZurek1982}
W.~K. Wootters and W.~H. Zurek, ``A single quantum cannot be cloned,'' {\em
  Nature}, vol.~299, p.~802, 1982.

\bibitem{BuzekHillery1996}
V.~Buzek and M.~Hillery, ``Quantum copying: Beyond the no-cloning theorem,''
  {\em Phys. Rev. A}, vol.~54, p.~1844, 1996.

\bibitem{GisinMassar1997}
N.~Gisin and S.~Massar, ``Opimal quantum cloning machines,'' {\em Phys. Rev.
  Lett.}, vol.~79, pp.~2153--2156, 1997.

\bibitem{Sorkin1987}
R.~Sorkin, ``A simplified derivation of stimulated emission by black holes,''
  {\em Class. Quant. Grav.}, vol.~5, pp.~L149--L155, 1987.

\bibitem{BirrellDavies1982}
N.~D. Birrell and P.~C.~W. Davies, {\em Quantum Fields in Curved Space}.
\newblock Cambridge: Cambridge U. Press, 1982.

\bibitem{CerfFiurasek2005}
N.~J. Cerf and J.~Fiurasek, ``Optical quantum cloning--{A} review,'' in {\em
  Progress in Optics} (E.~Wolf, ed.), vol.~49, Elsevier, 2006.

\bibitem{Scaranietal2005}
V.~Scarani, S.~Iblisdir, N.~Gisin, and A.~Acin, ``Quantum cloning,'' {\em Rev.
  Mod. Phys.}, vol.~77, pp.~1225--1256, 2005.

\bibitem{Brussetal1998}
D.~Bru\ss, D.~P. {Di Vincenzo}, A.~Ekert, C.~A. Fuchs, C.~Macchiavello, and
  J.~A. Smolin, ``Optimal universal and state-dependent quantum cloning,'' {\em
  Phys. Rev. A}, vol.~57, pp.~2368--2378, 1998.

\bibitem{Simon2000}
C.~Simon, G.~Weihs, and A.~Zeilinger, ``Optimal quantum cloning via stimulated
  emission,'' {\em Phys. Rev. Lett.}, vol.~84, pp.~2993--2996, 2000.

\bibitem{Fiurasek2001}
J.~Fiur{\'a}{\v s}ek, ``Optical implementation of continuous-variable quantum
  cloning machines,'' {\em Phys Rev Lett}, vol.~86, pp.~4942--4945, 2001.

\bibitem{Braunsteinetal2001}
S.~L. Braunstein, N.~J. Cerf, S.~Iblisdir, P.~van Loock, and S.~Massar,
  ``Optimal cloning of coherent states with a linear amplifier and beam
  splitters,'' {\em Phys. Rev. Lett.}, vol.~86, pp.~4938--4942, 2001.

\bibitem{Buzeketal1999}
V.~Buzek, M.~Hillery, and R.~F. Werner, ``Optimal manipulations with qubits:
  {U}niversal {NOT} gate,'' {\em Phys. Rev. A}, vol.~60, p.~R2626, 1999.

\bibitem{GisinPopescu1999}
N.~Gisin and S.~Popescu, ``Spin flips and quantum information for antiparallel
  spins,'' {\em Phys. Rev. Lett.}, vol.~83, p.~432, 1999.

\bibitem{MassarPopescu1995}
S.~Massar and S.~Popescu, ``Optimal extraction of information from finite
  quantum ensembles,'' {\em Phys. Rev. Lett.}, vol.~74, pp.~1259--1263, 1995.

\bibitem{LeonhardtNeumaier2004}
U.~Leonhardt and A.~Neumaier, ``Explicit effective hamiltonians from general
  linear-optical networks,'' {\em J. Optics B}, vol.~69, pp.~L1--L4, 2004.

\bibitem{Buzek1998}
V.~Buzek, D.~Kr\"ahmer, M.~Fontenelle, and W.~Schleich, ``Quantum statistics of
  grey-body radiation,'' {\em Phys. Lett. A}, vol.~239, pp.~1--5, 1998.

\bibitem{Schiffer1993}
M.~Schiffer, ``Is it possible to recover information from the black-hole
  radiation?,'' {\em Phys. Rev. D}, vol.~48, p.~1652, 1993.

\bibitem{Bradler2011}
K.~Br{\'a}dler, ``An infinite sequence of additive channels: the classical
  capacity of cloning channels,'' {\em IEEE Transactions on Information
  Theory}, vol.~57, no.~8, pp.~5497--5503, 2011.

\bibitem{BradlerAdami2014}
K.~Bradler and C.~Adami, ``The capacity of black holes to transmit quantum
  inforamtion,'' {\em J. High Energy Phys.}, vol.~1405, p.~095, 2014.
\newblock arXiv:1310.7914.

\bibitem{tHooft1985}
G.~'t~Hooft, ``On the quantum structure of a black hole,'' {\em Nuclear Physics
  B}, vol.~256, pp.~727--745, 1985.

\bibitem{Susskindetal1993}
L.~Susskind, L.~Thorlacius, and J.~Uglum, ``The stretched horizon and black
  hole complementarity,'' {\em Phys. Rev.}, vol.~48, no.~8, p.~3743, 1993.

\bibitem{Stephensetal1994}
C.~R. Stephens, G.~'t~Hooft, and B.~F. Whiting., ``Black-hole evaporation
  without information loss,'' {\em Classical and Quantum Gravity}, vol.~11,
  pp.~621--647, 1994.

\bibitem{Almheirietal2012}
A.~Almheiri, D.~Marolf, J.~Polchinski, and J.~Sully., ``Black holes:
  Complementarity or firewalls?.'' arXiv:1207.3123, 2012.

\bibitem{Braunsteinetal2013}
S.~L. Braunstein, S.~Pirandola, and K.~{\.Z}yczkowski, ``Better late than
  never: Information retrieval from black holes,'' {\em Physical Review
  Letters}, vol.~110, no.~10, p.~101301, 2013.

\bibitem{Davies1978}
P.~C.~W. Davies, ``Thermodynamics of black holes,'' {\em Reports Progress
  Physics}, vol.~41, pp.~1313--1355, 1978.

\end{thebibliography}

\end{document}